\documentclass[12pt]{article}

\usepackage{amssymb}
\usepackage{amsmath}
\usepackage{amscd}
\usepackage{latexsym}

\usepackage{cite}

\newcommand{\be}{\begin{equation}}
\newcommand{\ee}{\end{equation}}
\newcommand{\Dlt}{\Delta}
\newcommand{\dlt}{\delta}
\newcommand{\prt}{\partial}
\newcommand{\br}{{\bf r}}
\newcommand{\bk}{{\bf k}}

\newcommand{\bt}{\beta}
\newcommand{\vp}{\varphi}
\newcommand{\ep}{\varepsilon}
\newcommand{\al}{\alpha}
\newcommand{\ra}{\rightarrow}
\newcommand{\sgm}{\sigma}

\newcommand{\gm}{\gamma}
\newcommand{\om}{\omega}

\newcommand{\Gm}{\Gamma}
\newcommand{\dgr}{\dagger}
\newcommand{\lbd}{\lambda}
\newcommand{\Lbd}{\Lambda}

\newcommand{\rgl}{\rangle}
\newcommand{\lgl}{\langle}

\begin{document}

\begin{center}

{\Large{\bf Particle Fluctuations in Mesoscopic Bose Systems} \\ [5mm]

Vyacheslav I. Yukalov } \\ [3mm]

{\it Bogolubov Laboratory of Theoretical Physics, \\
Joint Institute for Nuclear Research, Dubna 141980, Russia \\ [2mm]
and \\ [2mm]
Instituto de Fisica de S\~ao Carlos, Universidade de S\~ao Paulo, \\
CP 369,  S\~ao Carlos 13560-970, S\~ao Paulo, Brazil  }
\end{center}

\vskip 2cm

\begin{abstract}
Particle fluctuations in mesoscopic Bose systems of arbitrary spatial dimensionality
are considered. Both ideal Bose gases and interacting Bose systems are studied in the
regions above the Bose-Einstein condensation temperature $T_c$ as well as below this
temperature. The strength of particle fluctuations defines whether the system is stable
or not. Stability conditions depend on the spatial dimensionality $d$ and on the
confining dimension $D$ of the system. The consideration shows that mesoscopic systems,
experiencing Bose-Einstein condensation, are stable when: (i) ideal Bose gas is confined
in a rectangular box of spatial dimension $d>2$ above $T_c$ and in a box of $d>4$ below
$T_c$; (ii) ideal Bose gas is confined in a power-law trap of a confining dimension
$D>2$ above $T_c$ and of a confining dimension $D>4$ below $T_c$; (iii) interacting Bose
system is confined in a rectangular box of dimension $d>2$ above $T_c$, while below
$T_c$ particle interactions stabilize the Bose-condensed system making it stable for
$d=3$; (iv) nonlocal interactions diminish the condensation temperature, as compared
with the fluctuations in a system with contact interactions.
\end{abstract}

\vskip 1.5cm
{\parindent=0pt
{\bf Keywords}: {\it Bose systems; asymptotic symmetry breaking; Bose-Einstein
condensation; particle fluctuations; stability of Bose systems} }

\vskip 2cm

\section{Introduction}

Theory of Bose systems has recently attracted high attention triggered by experimental
studies of cold trapped atoms (see, e.g., the books and review articles
\cite{Courteille_1,Andersen_2,Yukalov_3,Bongs_4,Yukalov_5,Lieb_6,Posazhennikova_7,
Morsch_8,Yukalov_9,Letokhov_10,Moseley_11,Bloch_12,Proukakis_13,Yurovsky_14,Pethick_15,
Yukalov_16,Yukalov_17,Yukalov_18,Yukalov_19}). A special attention has been payed to
the study of particle fluctuations, mainly considering three-dimensional macroscopic
Bose systems or harmonically trapped atoms. The importance of this problem has been
emphasized after the appearance of a number of papers claiming the occurrence of
thermodynamically anomalous particle fluctuations in the whole region below the condensation
temperature $T_c$ even for equilibrium three-dimensional interacting systems (a list of
the papers containing such claims has been summarized in \cite{Kocharovsky_20}). The origin
of the arising fictitious anomalies and the ways of avoiding them have been discussed in
detail in reviews \cite{Yukalov_16,Yukalov_17,Yukalov_18}.

It would not be strange if anomalously strong fluctuations would be found at the point
of a second-order phase transition. This would be natural, since at the point of a phase
transition the system is unstable and fluctuations in a system can drastically increase.
It is exactly the system instability that drives the phase transition and forces the
system to transfer to another state. But as soon as the transition to the other state
has happened, the real system becomes stable, and has to exhibit thermodynamically normal
fluctuations. It is therefore more than strange how thermodynamically anomalous
fluctuations could arise in realistic three-dimensional interacting systems.

Moreover, Bose-Einstein condensation is necessarily accompanied by the spontaneous
breaking of global gauge symmetry. From the mathematical point of view the similar
breaking of continuous symmetry occurs under magnetic phase transitions
\cite{Patashinsky_20}, hence, anomalous fluctuations of the order parameter should
appear in magnets below $T_c$. But thermodynamically anomalous fluctuations imply
the system instability \cite{Yukalov_20}. Therefore if such fluctuations would really arise
in the whole range below $T_c$, then neither superfluids nor magnets would exist. Fortunately,
it has been shown \cite{Yukalov_21,Yukalov_22} that thermodynamically anomalous
fluctuations in interacting three-dimensional equilibrium systems, discussed in
theoretical papers, are just calculational artifacts caused, briefly speaking, by the use
of a second-order approximation for calculating fourth-order terms.

The aim of the present paper is to extend the investigation of particle fluctuations
in Bose systems in several aspects: First, we consider mesoscopic systems that are
finite, although containing many particles $N\gg 1$. Taking into account a finite
number of particles requires to modify the definition of the Bose function by
introducing a finite cutoff responsible for the existence of a minimal wave vector
prescribed by the system geometry. Second, we analyze particle fluctuations above
as well as below $T_c$ for the Bose systems of arbitrary dimensionality, which allows
us to find the critical spatial dimension above which the system is stable. Third,
we consider two types of Bose systems, confined either in a rectangular box or in a
power-law trap. Fourth, the influence of nonlocal interactions on particle fluctuations
is analyzed, as compared to that of local interactions.

Throughout the paper, the system of units is employed where the Planck and Boltzmann
constants are set to one, $\hbar = 1$ and $k_B = 1$.

\section{Particle Fluctuations and Stability}

Here and in what follows, we consider {\it mesoscopic} systems that are finite, containing
a finite number of particles $N$, although with this number being rather large, $N \gg 1$.

Observable quantities are given by statistical averages $\langle \hat{A} \rangle$
of Hermitian operators $\hat{A}$. Fluctuations of the observable quantities are
characterized by the variance
$$
 {\rm var}(\hat A) \; \equiv \; \lgl \; \hat A^2 \; \rgl - \lgl\; \hat A \; \rgl^2 \;  .
$$
The observable is called {\it extensive}, when
\be
\label{1}
 \lgl\; \hat A \; \rgl  \; \propto \;  N \qquad ( N \gg 1) \;  ,
\ee
which is equivalent to the condition
\be
\label{2}
 \frac{\lgl\; \hat A \; \rgl}{N} \; \simeq \; const \qquad ( N \gg 1) \;  .
\ee

Fluctuations are termed {\it thermodynamically normal} if the inequalities
\be
\label{3}
0 \; \leq \; \frac{{\rm var}(\hat A)}{|\lgl\; \hat A \; \rgl |} \; < \; \infty
\ee
are valid for any $N$, which can also be represented as the condition
\be
\label{4}
 \frac{{\rm var}(\hat A)}{|\lgl\; \hat A \; \rgl |} \; \simeq \; const  \qquad
( N \gg 1) \; .
\ee
When these conditions do not hold, the fluctuations are called {\it thermodynamically
anomalous}. Sometimes, instead of the terms {\it thermodynamically normal} or
{\it thermodynamically anomalous}, one says, for short, that fluctuations are just
{\it normal} or {\it anomalous}.

Particle fluctuations, describing the fluctuations of the number of particles,
characterized by the number-of-particle operator $\hat{N}$, are quantified by the
relative variance
\be
\label{5}
  \frac{{\rm var}(\hat N)}{N} = \frac{1}{N} \; \left(
\lgl \; \hat N^2 \; \rgl - \lgl\; \hat N \; \rgl^2 \right) \; ,
\ee
where $N = \langle \hat{N} \rangle$. The fluctuations are normal when
\be
\label{6}
0 \; \leq \; \frac{{\rm var}(\hat N)}{N} \; < \; \infty
\ee
for any $N$, or in other words, when
\be
\label{7}
 \frac{{\rm var}(\hat N)}{N} \; \simeq \; const  \qquad ( N \gg 1) \;  .
\ee

The strength of particle fluctuations characterizes the system stability, since these
fluctuations are directly connected to the isothermal compressibility
\be
\label{8}
\kappa_T \equiv - \; \frac{1}{V} \left( \frac{\prt V}{\prt P} \right)_{TN} =
\frac{1}{\rho N} \left( \frac{\prt N}{\prt \mu} \right)_{TV}
\ee
by the equality
\be
\label{9}
 \kappa_T = \frac{{\rm var}(\hat N)}{\rho TN} \qquad
\left( \rho \equiv \frac{N}{V} \right) \; ,
\ee
with $\rho$ being average particle density. The system stability requires that
\be
\label{10}
  0 \; \leq \; \kappa_T \; < \; \infty
\ee
for any $N$, which yields conditions (\ref{6}) and (\ref{7}). The above relations
give us one of the ways for calculating the relative variance
\be
\label{11}
 \frac{{\rm var}(\hat N)}{N} = \rho T \kappa_T =
\frac{T}{N} \left( \frac{\prt N}{\prt \mu} \right)_{TV} \; .
\ee

\section{Ideal Gas in a Rectangular Box}

Bose systems in a rectangular box are not merely an interesting object allowing for
detailed calculations, but it can also be realized experimentally inside box-shaped
traps \cite{Gaunt_23,Navon_24,Lopes_25}.

\subsection{Modified Bose Function}

The grand Hamiltonian of a gas in a rectangular box of volume $V$ reads as
\be
\label{12}
 H = \hat H - \mu \hat N = \int \psi^\dgr(\br) \left( -\; \frac{\nabla^2}{2m}
- \mu\right) \psi(\br) \; d\br \; ,
\ee
where the integration is over the given volume $V$. Assuming periodic continuation
of the box, the field operators can be expanded in plane waves,
\be
\label{13}
 \psi(\br) = \sum_k a_k \vp_k(\br) \; , \qquad
\vp_k(\br) = \frac{1}{\sqrt{V}} \; e^{i\bk\cdot\br} \;  ,
\ee
which gives
\be
\label{14}
 H = \sum_k \om_k a_k^\dgr a_k \qquad
\left( \om_k = \frac{k^2}{2m} - \mu \right) \; .
\ee

The total number of particles is the sum
\be
\label{15}
 N = N_0 + N_1 \; , \qquad N_1 = \sum_{k\neq k_0} n_k \;  ,
\ee
where $N_0$ is the number of condensed particles, while $N_1$ is the number of
uncondensed particles, with the momentum distribution
\be
\label{16}
 n_k \equiv \lgl \; a_k^\dgr a_k \; \rgl =
\left( e^{\bt\om_k} - 1 \right)^{-1} \;  .
\ee
Here $\beta = 1/T$ is inverse temperature. For a large number of particles, the
sums over momenta can be represented as the integrals,
$$
 \sum_k n_k \ra V \int n_k \; \frac{d\bk}{(2\pi)^d} \;  ,
$$
where $d$ is spatial dimensionality. In the case of isotropic functions under the
integrals, it is possible to pass to spherical coordinates. However, it is necessary
to take into account that for a finite system the values of the wave vectors start
not from zero but from a finite minimal momentum $k_0$ that can be estimated as
\be
\label{17}
 k_0 = \frac{2\pi}{L} = \frac{2\pi}{aN^{1/d}} \;  ,
\ee
with the box volume
$$
 V = L^d \; , \qquad L = a N^{1/d} \;  ,
$$
where $a$ is mean interparticle distance. Thus the integration over the momenta
takes the form
\be
\label{18}
 \int \frac{d\bk}{(2\pi)^d} \; \ra \;
\frac{2}{(4\pi)^{d/2}\Gm(d/2)} \int_{k_0}^\infty k^{d-1}\; dk \; ,
\ee
where the lower limit is given by the cutoff prescribed by the minimal quantity $k_0$.
Then the number of uncondensed particles becomes proportional to the {\it modified Bose
function}
\be
\label{19}
 g_n(z) \equiv
\frac{1}{\Gm(n)} \int_{u_0}^\infty \frac{z u^{n-1}}{e^u-z} \; du \; ,
\ee
with $z \equiv \exp(\beta \mu)$ being the fugacity and where the lower limit is given
by the cutoff
\be
\label{20}
 u_0   = \frac{k_0^2}{2mT} = \frac{\ep_0}{T}
\ee
defined by the minimal energy
\be
\label{21}
 \ep_0 = \frac{2\pi^2}{ma^2}\; N^{-2/d} \;  .
\ee
Since the minimal energy (\ref{21}) tends to zero for large $N$, it is admissible to
keep in mind that
\be
\label{22}
 u_0 \; \ll \; 1 \qquad ( N \gg 1 ) \;  .
\ee
In this way, the relative variance (\ref{11}) can be expressed through the derivative
of the modified Bose function (\ref{19}). The latter differs from the standard Bose function
by the existence of a nonzero lower integration limit defined by the minimal wave vector.

\subsection{Fluctuations above Condensation Temperature}

At temperatures above the condensation point, there are no condensed particles, so that
the total number of particles reads as
\be
\label{23}
 N = \frac{V}{\lbd_T^d}\; g_{d/2}(z) \qquad ( T \geq T_c) \;  ,
\ee
where
$$
 \lbd_T \equiv \sqrt{ \frac{2\pi}{mT} }
$$
is thermal wavelength. Hence the relative variance (\ref{11}) is
\be
\label{24}
\frac{{\rm var}(\hat N)}{N} = \frac{z}{\rho\lbd_T^d} \;
\frac{\prt g_{d/2}(z)}{\prt z}  \qquad ( T > T_c ) \; ,
\ee
where $\rho \equiv N/V$ is particle density.

Estimating the Bose function above $T_c$, where $z < 1$, we find
\be
\label{25}
 g_n(z) = - \; \frac{z}{(1-z)\Gm(1+n)}\left[ u_0^n - \;
\frac{nu_0^{1+n}}{(1+n)(1-z)} \right] \qquad ( n < 0 \; , ~ z < 1 ) \;  .
\ee
In particular,
\be
\label{26}
  g_{-1/2}(z) = - \; \frac{z}{\sqrt{\pi}(1-z)}\left( u_0^{-1/2} +
\frac{u_0^{1/2}}{1-z} \right) \qquad ( z < 1 )
\ee
and
\be
\label{27}
  g_0(z) = -\; \frac{z}{1-z} \qquad ( z < 1 ) \; .
\ee

Calculating the derivatives of the modified Bose functions requires to be
attentive, since some of the derivatives are different from those for the
standard Bose functions. Generally, we have
\be
\label{28}
 \frac{\prt g_n(z)}{\prt z} = \frac{1}{z} \; g_{n-1}(z) +
\frac{u_0^{n-1}}{\Gm(n)(1-z+u_0)} \;  .
\ee
Using the smallness of $u_0$, we can write
$$
 \frac{u_0^{n-1}}{1-z+u_0} \; \simeq \; \frac{1}{1-z} \left( u_0^{n-1} - \;
\frac{u_0^n}{1-z} \right) \qquad ( z < 1 ) \;  .
$$
Therefore for $n < 1$, we find
\be
\label{29}
 \frac{\prt g_n(z)}{\prt z} = -\; \frac{u_0^n}{(1-z)^2\Gm(1+n)} \qquad
(n < 1 \; , ~ z < 1 ) \;  .
\ee
While for $n > 1$, keeping the main terms, we get
\be
\label{30}
 \frac{\prt g_n(z)}{\prt z} = \frac{1}{z} \; g_{n-1}(z) \qquad
(n > 1 \; , ~ z < 1 ) \;   .
\ee
We shall also need the derivatives
$$
 \frac{\prt g_{1/2}(z)}{\prt z} = -\; \frac{2u_0^{1/2}}{\sqrt{\pi}(1-z)^2}
\qquad  ( d = 1 \; , ~ z < 1 ) \; ,
$$
$$
 \frac{\prt g_1(z)}{\prt z} = -\; \frac{u_0}{(1-z)^2} \qquad
( d = 2 \; , ~ z < 1 ) \; ,
$$
$$
\frac{\prt g_{3/2}(z)}{\prt z} = \frac{1}{z} \; g_{1/2}(z) \qquad
( d = 3 \; , ~ z < 1 ) \; .
$$

For the relative variance, depending on the space dimensionality, we obtain
$$
\frac{{\rm var}(\hat N)}{N} = -\; \frac{2z}{\sqrt{\pi}(1-z)^2\rho\lbd_T}
\left( \frac{\ep_0}{T}\right)^{1/2} \qquad ( d = 1 \; , ~ T > T_c ) \; ,
$$
$$
\frac{{\rm var}(\hat N)}{N} = -\; \frac{z}{(1-z)^2\rho\lbd_T^2}
\left( \frac{\ep_0}{T}\right) \qquad ( d = 2 \; , ~ T > T_c ) \; ,
$$
\be
\label{31}
\frac{{\rm var}(\hat N)}{N} =  \frac{1}{\rho\lbd_T^3} \; g_{1/2}(z) \qquad
( d = 3 \; , ~ T > T_c ) \; .
\ee
The negative values of the variance for $d = 1$ and $d = 2$ show that these
low-dimensional systems are unstable. But the gas is stable in three dimensions.

As follows from the derivative
\be
\label{32}
 \frac{\prt g_{d/2}(z)}{\prt z} = \frac{1}{z} \; g_{(d-2)/2}(z) \qquad
 ( d > 2 \; , ~ z < 1 ) \; ,
\ee
the system is stable for $d > 2$. That is, the critical spatial dimension, above
which the uncondensed gas in a rectangular box is stable is $d_c = 2$, so that the
stability condition is
\be
\label{33}
 d \; > \; d_c = 2 \qquad ( T > T_c ) \;  .
\ee

\subsection{Condensation Temperature of a Gas in a Rectangular Box}

At the temperature of Bose condensation, the chemical potential becomes zero,
$\mu = 0$, because of which $z = 1$. The total number of particles
\be
\label{34}
N = \frac{V}{\lbd_T^d} \; g_{d/2}(1) \qquad ( T = T_c )
\ee
defines the critical temperature
\be
\label{35}
 T_c = \frac{2\pi}{m} \left[ \frac{\rho}{g_{d/2}(1)}\right]^{2/d} \;  .
\ee

For different dimensionalities, we have
$$
g_{1/2}(1) = \frac{2}{\sqrt{\pi}} \; u_0^{-1/2} \qquad ( d = 1 ) \; ,
$$
$$
g_1(1) = -\ln u_0  \qquad ( d = 2 ) \; ,
$$
$$
g_{3/2}(1) = \zeta(3/2)  \qquad ( d = 3 ) \; .
$$
This gives the critical temperatures for a one-dimensional system
\be
\label{36}
 T_c = \frac{\pi\rho}{\sqrt{2m}} \; \ep_0^{1/2}   \qquad ( d = 1 ) \; ,
\ee
and for a two-dimensional system
\be
\label{37}
  T_c = \frac{2\pi\rho}{m\ln(T_c/\ep_0)}   \qquad ( d = 2 ) \;  .
\ee
Iterating the latter equation and taking into account that
$$
 \frac{T_c}{\ep_0} \; \ll \; \exp\left( \frac{2\pi\rho}{m\ep_0}\right) \; ,
\qquad
\ep_0 \; \propto \; N^{-2/d} \qquad ( N \gg 1 ) \;  ,
$$
we obtain
\be
\label{38}
  T_c = \frac{2\pi\rho}{m\ln(2\pi\rho/m\ep_0)}   \qquad ( d = 2 ) \;   .
\ee
For a three-dimensional box, we have the known result
\be
\label{39}
  T_c = \frac{2\pi}{m} \left[ \frac{\rho}{\zeta(3/2)} \right]^{2/3} \qquad
( d = 3 ) \;  .
\ee

The critical temperatures at large $N \gg 1$ scale as
$$
T_c \; \propto \; \frac{1}{N} \qquad ( d = 1 ) \; ,
$$
$$
T_c \; \propto \; \frac{1}{\ln N} \qquad ( d = 2 ) \; ,
$$
\be
\label{40}
T_c \; \propto \; const \qquad ( d = 3 ) \;   .
\ee
For $d \leq 2$ the critical temperature diminishes to zero as $N$ increases. It
remains finite for $d > 2$. Recall that, as is found above, the system is unstable
for $d \leq 2$. Thus the Bose gas in a box is stable in the case where the critical
temperature remains finite for large $N$.

\subsection{Fluctuations below Critical Temperature}

Below the critical temperature, there appears Bose-Einstein condensate, so that the
number-of-particle operator becomes the sum of the number of condensed particles
$N_0$ and the number-of-particle operator $\hat{N}_1$ of uncondensed particles,
\be
\label{41}
 \hat N = N_0 + \hat N_1 \qquad ( T < T_c ) \;  .
\ee
When the condensate function $\eta$ is introduced by means of the Bogolubov shift
\cite{Bogolubov_26,Bogolubov_27,Bogolubov_28} of the field operator
$$
 \psi(\br) \ra \eta(\br) + \psi_1(\br) \;  ,
$$
then particle fluctuations are defined by the fluctuations of uncondensed particles
(see detailed explanations in Refs.
\cite{Yukalov_3,Yukalov_9,Yukalov_16,Yukalov_17,Yukalov_18}),
$$
 {\rm var}(\hat N) = {\rm var}(\hat N_1) \;  .
$$
The average number of uncondensed particles is
\be
\label{42}
 N_1 = \frac{V}{\lbd_T^d} \; g_{d/2}(1) \qquad ( T < T_c ) \;  .
\ee
So that the relative particle variance reads as
\be
\label{43}
 \frac{{\rm var}(\hat N)}{N} = \frac{1}{\rho\lbd_T^d} \;
\lim_{z\ra 1} \frac{\prt g_{d/2}(z)}{\prt z}     \qquad ( T < T_c ) \;  .
\ee

For a mesoscopic system, we have
\be
\label{44}
 \lim_{z\ra 1} \frac{\prt g_n(z)}{\prt z} =
g_{n-1}(1) + \frac{u_0^{n-2}}{\Gm(n)} \;  .
\ee
Notice that the last term here would be absent for a macroscopic system. In
particular,
$$
\lim_{z\ra 1} \frac{\prt g_{1/2}(z)}{\prt z} =  g_{-1/2}(1) +
\frac{u_0^{-3/2}}{\sqrt{\pi}} \;   .
$$
Since
$$
g_{-1/2}(1) = - \; \frac{1}{3\sqrt{\pi}} \; u_0^{-3/2} \; ,
$$
we find
$$
 \lim_{z\ra 1} \frac{\prt g_{1/2}(z)}{\prt z} =
\frac{2}{3\sqrt{\pi}} \; u_0^{-3/2} \;   .
$$
We also need the limits
$$
\lim_{z\ra 1} \frac{\prt g_1(z)}{\prt z} =  \frac{1}{u_0} \;  ,
$$
and
$$
 \lim_{z\ra 1} \frac{\prt g_{3/2}(z)}{\prt z} =  g_{1/2}(1) +
\frac{2}{\sqrt{\pi}} \; u_0^{-1/2} \;  .
$$
Using $g_{1/2}(1)$ from the previous subsection, we get
$$
\lim_{z\ra 1} \frac{\prt g_{3/2}(z)}{\prt z} =
\frac{4}{\sqrt{\pi}}\; u_0^{-1/2} \; .
$$

Generally, from the expression
\be
\label{45}
\lim_{z\ra 1} \frac{\prt g_{d/2}(z)}{\prt z} =  g_{(d-2)/2}(1) +
\frac{u_0^{(d-4)/2}}{\Gm(d/2)}
\ee
we see that the last term here increases with $N$ by a power law, when $d<4$,
while it increases logarithmically for $d = 4$,
$$
\lim_{z\ra 1} \frac{\prt g_2(z)}{\prt z} = 1 - \ln u_0
\qquad ( d = 4 ) \;  .
$$

For the relative variance (\ref{43}) we obtain
$$
\frac{{\rm var}(\hat N)}{N} =
\frac{2}{3\sqrt{\pi}\rho\lbd_T} \left( \frac{T}{\ep_0} \right)^{3/2}
\qquad ( d = 1 ) \; ,
$$
$$
\frac{{\rm var}(\hat N)}{N} =
\frac{1}{\rho\lbd_T^2} \left( \frac{T}{\ep_0} \right)
\qquad ( d = 2 ) \; ,
$$
$$
\frac{{\rm var}(\hat N)}{N} =
\frac{4}{\sqrt{\pi}\rho\lbd_T^3} \left( \frac{T}{\ep_0} \right)^{1/2}
\qquad ( d = 3 ) \; ,
$$
\be
\label{46}
\frac{{\rm var}(\hat N)}{N} =
\frac{1}{\rho\lbd_T^4} \left( \frac{T}{\ep_0} \right)
\qquad ( d = 4 ) \;   .
\ee
Keeping in mind that $\varepsilon_0 \propto N^{-2/d}$, the scaling of these
expressions with respect to $N$ is as follows:
$$
\frac{{\rm var}(\hat N)}{N} \; \propto \; N^3   \qquad ( d = 1 ) \; ,
$$
$$
\frac{{\rm var}(\hat N)}{N} \; \propto \; N   \qquad ( d = 2 ) \; ,
$$
$$
\frac{{\rm var}(\hat N)}{N} \; \propto \; N^{1/3}   \qquad ( d = 3 ) \; ,
$$
\be
\label{47}
\frac{{\rm var}(\hat N)}{N} \; \propto \; \ln N   \qquad ( d = 4 ) \;   .
\ee
This shows that for all dimensions below and including $4$ particle fluctuations
are anomalous, corresponding to unstable systems. In that sense, the dimension $4$
is critical, implying that the stability condition for a condensed gas in a box is
\be
\label{48}
 d \; > \; d_c = 4 \qquad ( T < T_c ) \;  .
\ee

\section{Ideal Gas in a Power-Law Trap}

Power-law traps are the most often used devices for trapping particles. Here we
study particle fluctuations and the related stability of mesoscopic clouds in such
traps.

\subsection{Modified Semiclassical Approximation}

The general form of confining potentials, employed in power-law traps, can be
represented as
\be
\label{49}
 U(\br) = \sum_{\al=1}^d \frac{\om_\al}{2} \;
\left| \; \frac{r_\al}{l_\al} \;\right|^{n_\al} \;  ,
\ee
where
$$
l_\al \equiv \frac{1}{\sqrt{m\om_\al} }
$$
is the effective trap radius in the $\alpha$ direction. As a whole, a trap can be
characterized by the effective trap frequency $\omega_0$ and effective length $l_0$
connected by the relations
\be
\label{50}
\om_0 \equiv \left( \prod_{\al=1}^d \om_\al \right)^{1/d} =
\frac{1}{ml_0^2} \; ,
\qquad
l_0 \equiv \left( \prod_{\al=1}^d l_\al \right)^{1/d} =
\frac{1}{\sqrt{m\om_0}} \;  .
\ee
In the limit $n_\alpha \ra \infty$, we return to a rectangular box.

When the effective trap frequency is much lower than temperature,
\be
\label{51}
 \frac{\om_0}{T} \ll 1 \;  ,
\ee
it is possible to resort to the semiclassical approximation that, however, needs
to be modified for considering mesoscopic systems \cite{Yukalov_18,Yukalov_29}.

In the semiclassical approximation, one defines the density of states
$$
 \rho(\ep) = \frac{(2m)^{d/2}}{(4\pi)^{d/2}\Gm(d/2)}
\int_{\mathbb{V}_\ep} [ \ep - U(\br) ]^{d/2-1} \; d\br \;  ,
$$
in which
$$
\mathbb{V}_\ep \equiv \{ \br: \; U(\br) \leq \ep \}
$$
is the volume available for particle motion.

For trapped particles, an important notion is the {\it confining dimension}
\cite{Yukalov_18,Yukalov_29}
\be
\label{52}
D \equiv d + \sum_{\al=1}^d \frac{2}{n_\al} \; .
\ee

The density of states for the power-law potential (\ref{49}) reduces to
\be
\label{53}
 \rho(\ep) = \frac{\ep^{D/2-1}}{\gm_D\Gm(D/2)} \;  ,
\ee
where we use the notation
$$
 \gm_D \; \equiv \; \frac{\pi^{d/2}}{2^{D/2}}
\prod_{\al=1}^d \; \frac{\om_\al^{1/2+1/n_\al}}{\Gm(1+1/n_\al)} \;  .
$$

In the normal state above $T_c$, the number of particles is given by the formula
\be
\label{54}
 N = \frac{T^{D/2}}{\gm_D} \; g_{D/2}(z) \qquad ( T \geq T_c ) \; .
\ee
And we again meet the Bose function that has to be modified according to definition
(\ref{19}) by using the integral cutoff
\be
\label{55}
u_0 = \frac{\ep_0}{T} \qquad ( \ep_0 \sim \om_0 ) \;   ,
\ee
with $\varepsilon_0$ being the lowest energy level in the trap, which is of order
of $\om_0$.

\subsection{Condensation Temperature of a Gas in a Power-Law Trap}

At the critical temperature $T_c$, we have $\mu=0$ and $z=1$. Then equation
(\ref{54}) yields
\be
\label{56}
  T_c = \left[ \frac{\gm_D N}{g_{D/2}(1)}\right]^{2/D} \; .
\ee
The modified Bose function, depending on the confining dimension, takes the forms
$$
g_{D/2}(1) = \frac{2}{(2-D)\Gm(D/2)}\left( \frac{T}{\ep_0}\right)^{1-D/2}
\qquad ( D < 2)  \; ,
$$
$$
g_1(1) = \ln\; \frac{T}{\ep_0} \qquad ( D = 2 ) \; ,
$$
$$
g_{D/2}(1) = \zeta \left( \frac{D}{2} \right)       \qquad ( D > 2)  \; .
$$

If $D < 2$, the spatial dimension can only be $d = 1$, when
$$
\gm_D = \frac{\sqrt{\pi}}{\Gm(D)} \; \left( \frac{\om_0}{2} \right)^{D/2}
\qquad ( d = 1)  \;   .
$$
Then the critical temperature is
\be
\label{57}
T_c = \frac{\sqrt{\pi}}{\Gm(D)} \left( 1 - \;\frac{D}{2}\right)
\Gm\left( \frac{D}{2} \right) \left(  \frac{\om_0}{2\ep_0}\right)^{D/2} N \ep_0
\qquad ( D < 2 \; , ~ d = 1) \;   .
\ee

The confining dimension equals two, $D = 2$, when $d = 1$ and $n = 2$, so that
$$
\gm_2 = \om_0 \qquad ( D = 2 \; , ~ d = 1 \; , ~ n = 2 ) \; .
$$
This yields the critical temperature
\be
\label{58}
 T_c = \frac{N\om_0}{\ln(T_c/\ep_0)} \qquad ( D = 2 \; , ~ d = 1 ) \; .
\ee
For large $N$, one has
$$
 \frac{T_c}{\ep_0} \; \ll \; \exp\left( \frac{\om_0}{\ep_0}\; N \right) \;  ,
$$
since for a one-dimensional harmonic oscillator $\varepsilon_0 = \omega_0/2$.
Because of this, the critical temperature (\ref{58}) can be simplified to
\be
\label{59}
 T_c = \frac{N\om_0}{\ln(2N)} \;  .
\ee

And for the confining dimension larger than two, the critical temperature is
\be
\label{60}
 T_c = \left[ \frac{\gm_DN}{\zeta(D/2)}\right]^{2/D}   \qquad ( D > 2 ) \;  .
\ee

In the case of harmonic traps, when $n_\al=2$, hence $D=2d$ and $\gm_D=\om_0^d$,
the critical temperature becomes
$$
T_c = \left[ \frac{N}{\zeta(d)}\right]^{1/d} \om_0  \qquad
( n_\al = 2\; , ~ d > 1 ) \;   .
$$

\subsection{Scaling with Respect to Particle Number}

As is explained in Sec. 2, extensive observables are proportional to the number
of particles $N$, when this number is large. This definition prescribes the
scaling of the system characteristics with respect to $N$. As a representative
of an observable quantity, we may take, e.g., internal energy
\be
\label{61}
  \lgl \; \hat H \; \rgl = \lgl \;  H \; \rgl  + \mu N \; .
\ee
This is an extensive quantity satisfying the condition
\be
\label{62}
 \frac{\lgl \; \hat H \; \rgl}{N} \; \simeq \; const \qquad ( N \gg 1 ) \;  .
\ee
For the considered case of a gas in a power-law trap, we have
\be
\label{63}
\frac{\lgl \; \hat H \; \rgl}{N} =
\frac{D g_{1+D/2}(z)}{2N\gm_D} \; T^{1+D/2} \; .
\ee
The function $g_{1+D/2}(z)$ is finite for all $D>0$ and all $z$. Hence condition
(\ref{62}) implies
\be
\label{64}
 N\gm_D \; \simeq \; const \qquad ( N \gg 1)  ; .
\ee

To make the consideration slightly less cumbersome, let us set the powers $n_\al=n$
for the trapping potential. Then the confining dimension is
\be
\label{65}
 D = \left( 1 + \frac{2}{n}\right) d \;  .
\ee
And $\gamma_D$ becomes
$$
\gm_D = \frac{\pi^{d/2}}{\Gm^d(1+1/n)} \left( \frac{\om_0}{2} \right)^{D/2} \;   ,
$$
which tells us that
$$
 \gm_D \; \propto \; \om_0^{D/2} \qquad ( N \gg 1) \;  .
$$
Therefore, $\omega_0$ scales as
\be
\label{66}
\om_0 \; \propto \; \frac{1}{N^{2/D}}  \qquad ( N \gg 1)  \;  .
\ee

Using this scaling and the fact that $\om_0\sim \ep_0$, we see that the critical
temperatures from the previous subsection behave as
$$
T_c \; \propto \; \frac{1}{N^{2/D-1}} \qquad ( D < 2 ) \; ,
$$
$$
T_c \; \propto \; \frac{1}{\ln N} \qquad ( D = 2 ) \; ,
$$
\be
\label{67}
 T_c \; \propto \; const \qquad ( D > 2 ) \;  .
\ee

\subsection{Fluctuations above Condensation Temperature}

Particle fluctuations above the condensation temperature are described by the
formula
\be
\label{68}
 \frac{{\rm var}(\hat N)}{N} =
\frac{T^{D/2}}{N\gm_D} \; z \; \frac{\prt g_{D/2}(z)}{\prt z}
\qquad ( T > T_c ) \;  ,
\ee
where $z < 1$. For the modified Bose function, we have
\be
\label{69}
\frac{\prt g_m(z)}{\prt z} = \frac{1}{z} \; g_{m-1}(z) +
\frac{1}{(1-z)\Gm(m)} \left( u_0^{m-1} - \; \frac{u_0^m}{1-z}\right )
\qquad ( z < 1) \; ,
\ee
with the value
\be
\label{70}
 g_{m-1}(z) = -\; \frac{z}{(1-z)\Gm(m)}
\left[ u_0^{m-1} - \; \frac{m-1}{m(1-z)} \; u_0^m \right ]
\qquad (m < 1 \; , ~ z < 1)
\ee
for $m < 1$. Summarizing, we have the derivatives
$$
\frac{\prt g_m(z)}{\prt z} = - \; \frac{u_0^m}{(1-z)^2\Gm(1+m)}
\qquad (m < 1 \; , ~ z < 1) \; ,
$$
$$
\frac{\prt g_1(z)}{\prt z} = - \; \frac{u_0}{(1-z)^2}
\qquad (m = 1 \; , ~ z < 1) \; ,
$$
$$
\frac{\prt g_m(z)}{\prt z} = \frac{1}{z} \; g_{m-1}(z)
\qquad (m > 1 \; , ~ z < 1) \;  .
$$

From here we find the relative variance
$$
\frac{{\rm var}(\hat N)}{N} = - \;
\frac{z T^{D/2}}{(1-z)^2 N\gm_D\Gm(1+D/2)} \left( \frac{\ep_0}{T}\right)^{D/2}
\qquad (D < 2 \; , ~ T > T_c ) \; ,
$$
$$
\frac{{\rm var}(\hat N)}{N} = - \;
\frac{z T}{(1-z)^2 N\gm_2} \left( \frac{\ep_0}{T}\right)
\qquad (D = 2 \; , ~ T > T_c ) \; ,
$$
\be
\label{71}
\frac{{\rm var}(\hat N)}{N} = \frac{T^{D/2}}{N\gm_D} \; g_{D/2-1}(z)
\qquad (D > 2 \; , ~ T > T_c ) \;  ,
\ee
characterizing particle fluctuations above the critical temperature. For $D\leq 2$,
the variance is negative, which means instability. The system is stable only for
$D > 2$, giving the stability condition
\be
\label{72}
 d + \sum_{\al=1}^d \frac{2}{n_\al} \; > \; 2 \qquad ( T > T_c ) \;  .
\ee

\subsection{Fluctuations below Condensation Temperature}

Below the condensation temperature, where $\mu = 0$ and $z = 1$, the number of
uncondensed particles reads as
\be
\label{73}
N_1 = \frac{T^{D/2}}{\gm_D}\; g_{D/2}(1) \qquad ( T \leq T_c ) \; .
\ee
The variance of the total number of particles coincides with that of the uncondensed
particles, which leads to
\be
\label{74}
 \frac{{\rm var}(\hat N)}{N} = \frac{T^{D/2}}{N\gm_D} \;
\lim_{z\ra 1} \; \frac{\prt g_{D/2}(z)}{\prt z}    \qquad ( T < T_c ) \;  .
\ee
For the derivative in the right-hand side of the above formula, we have
\be
\label{75}
\lim_{z\ra 1} \; \frac{\prt g_{D/2}(z)}{\prt z} = g_{D/2-1}(1) +
\frac{u_0^{D/2-2}}{\Gm(D/2)} \; .
\ee
Employing the values
$$
g_{D/2-1}(1) = \frac{2}{(4-D)\Gm(D/2-1)}\; u_0^{D/2-2}   \qquad ( D < 4 ) \; ,
$$
$$
 g_1(1) = -\ln u_0 \qquad ( D = 4 ) \;  ,
$$
we get the derivatives
$$
\lim_{z\ra 1} \; \frac{\prt g_{D/2}(z)}{\prt z} = \left[ \frac{2}{(4-D)\Gm(D/2-1)}
+ \frac{1}{\Gm(D/2)} \right] u_0^{D/2-2}    \qquad ( D < 4 ) \; ,
$$
$$
\lim_{z\ra 1} \; \frac{\prt g_2(z)}{\prt z} = -\ln u_0 \qquad ( D = 4 ) \;  ,
$$
$$
 \lim_{z\ra 1} \; \frac{\prt g_{D/2}(z)}{\prt z} = g_{D/2-1}(1) =
\zeta\left( \frac{D}{2} - 1\right)  \qquad ( D > 4 ) \; .
$$
In that way, we come to the relative variances
$$
\frac{{\rm var}(\hat N)}{N} = \frac{T^{D/2}}{N\gm_D}
\left[ \frac{2}{(4-D)\Gm(D/2-1)} + \frac{1}{\Gm(D/2)} \right]
\left( \frac{T}{\ep_0}\right)^{2-D/2}    \qquad ( D < 4 ) \; ,
$$
$$
\frac{{\rm var}(\hat N)}{N} = \frac{T^2}{N\gm_4}\; \ln \left( \frac{T}{\ep_0}\right)
\qquad ( D = 4 ) \; ,
$$
\be
\label{76}
 \frac{{\rm var}(\hat N)}{N} = \frac{T^{D/2}}{N\gm_D}\;
\zeta \left( \frac{D}{2} - 1 \right)   \qquad ( D > 4 ) \; .
\ee
Keeping in mind that $\varepsilon_0 \propto N^{-2/D}$ results in the scaling
$$
\frac{{\rm var}(\hat N)}{N} \; \propto \; N^{(4-D)/D}  \qquad ( D < 4 ) \; ,
$$
$$
\frac{{\rm var}(\hat N)}{N} \; \propto \; \ln N  \qquad ( D = 4 ) \; ,
$$
\be
\label{77}
 \frac{{\rm var}(\hat N)}{N} \; \propto \; const  \qquad ( D > 4 ) \;  .
\ee
This tells us that the system is stable only for $D > 4$. So that the stability
condition is
\be
\label{78}
 d + \sum_{\al=1}^d \frac{2}{n_\al} \; > \; 4 \qquad ( T < T_c ) \;  .
\ee

Notice that in the case of the often considered harmonic potential, when
$n_\al=2$, we have $D=2d$ and $\gm_D=\om_0^d$. Then the stability condition
(\ref{78}) reduces to the condition $d>2$. And the relative particle variance
reads as
$$
\frac{{\rm var}(\hat N)}{N} = \frac{\zeta(d-1)}{\zeta(d)} \;
\left( \frac{T}{T_c} \right)^d  \qquad ( n_\al = 2 \; , ~ d > 2 ) \; .
$$

\section{Interacting Bose System above Condensation Temperature}

The grand Hamiltonian for a system of interacting Bose particles is
$$
 H = \int \psi^\dgr(\br) \left( -\; \frac{\nabla^2}{2m} - \mu
\right) \psi(\br) \; d\br \; +
$$
\be
\label{79}
+ \; \frac{1}{2} \int \psi^\dgr(\br) \psi^\dgr(\br')
\Phi(\br-\br') \psi(\br') \psi(\br)  \; d\br d\br' \; .
\ee
For generality, we consider a nonlocal isotropic interaction potential
$\Phi(\br)=\Phi(r)$, where $r\equiv |\br|$. The integration is assumed to be over
a rectangular box of volume $V$ confining the system.

In the Hartree-Fock approximation, the Hamiltonian takes the form
$$
 H_{HF} = E_{HF} + \int \psi^\dgr(\br) \left( -\; \frac{\nabla^2}{2m} - \mu
\right) \psi(\br) \; d\br \; +
$$
\be
\label{80}
 + \;
\int \Phi(\br-\br') \left[ \rho(\br') \psi^\dgr(\br) \psi(\br) +
\rho(\br',\br) \psi^\dgr(\br') \psi(\br) \right] \; d\br d\br' \; ,
\ee
where
$$
E_{HF} = - \;  \frac{1}{2} \int \Phi(\br-\br') \left[ \rho(\br)\rho(\br') +
|\; \rho(\br,\br')\; |^2 \right] \; d\br d\br'
$$
and the notations are used for the single-particle density matrix
\be
\label{81}
\rho(\br,\br') = \lgl \; \psi^\dgr(\br') \psi(\br) \; \rgl
\ee
and the particle density
\be
\label{82}
 \rho(\br) =  \rho(\br,\br) = \lgl \; \psi^\dgr(\br) \psi(\br) \; \rgl  \; .
\ee

Employing the expansion of the field operators over plane waves, as in equation
(\ref{13}), we get the Hamiltonian
\be
\label{83}
 H_{HF} = E_{HF} + \sum_k \om_k a_k^\dgr a_k \;  ,
\ee
in which
$$
  E_{HF} = - \;  \frac{1}{2} \; \rho \Phi_0 N - \;
\frac{1}{2V} \sum_{kp} n_k n_p \Phi_{k+p} \; ,
$$
$\Phi_k$ is a Fourier transform of $\Phi({\bf r})$, and
\be
\label{84}
 \Phi_0 = \int \Phi(\br) \; d\br \;  .
\ee
The momentum distribution is given by expression (\ref{16}), with the spectrum
\be
\label{85}
 \om_k = \frac{k^2}{2m} + \rho \Phi_0 +
\frac{1}{V} \sum_p n_p \Phi_{k+p} - \mu \;  .
\ee
The function $n_p$ possesses a maximum at $p \ra 0$, because of which it is possible
to use the approximation \cite{Yukalov_30,Yukalov_31}
\be
\label{86}
 \sum_p \; n_p \Phi_{k+p} \cong \Phi_k \sum_p n_p
\ee
giving
\be
\label{87}
  \om_k = \frac{k^2}{2m} + \rho ( \Phi_0 + \Phi_k ) - \mu \;  .
\ee
Introducing the effective interaction radius by the relation
\be
\label{88}
r_{eff}^2 \equiv \frac{\int \Phi(\br) r^2d\br}{\int \Phi(\br)d\br} =
\frac{4\pi}{\Phi_0} \int_0^\infty \Phi(\br) r^4 \; d\br
\ee
shows that the long-wave limit of $\Phi_k$ is
\be
\label{89}
 \Phi_k \simeq \left( 1 - \; \frac{1}{6}\; k^2 r^2_{eff}\right) \Phi_0 \;  .
\ee
Then spectrum (\ref{87}) can be represented as
\be
\label{90}
  \om_k \; \simeq \; \frac{k^2}{2m^*} \; - \; \mu_{eff} \qquad ( k \ra 0 ) \; ,
\ee
with the effective mass
\be
\label{91}
m^* \equiv \frac{m}{1-\rho\Phi_0 r^2_{eff}/3}
\ee
and effective chemical potential
\be
\label{92}
  \mu_{eff} \equiv \mu - 2\rho \Phi_0 \; .
\ee

In this approximation, the number of particles acquires the same form (\ref{23}),
however with the notations
\be
\label{93}
 \lbd_T \equiv \sqrt{\frac{2\pi}{m^*T} } \; , \qquad z \equiv
\exp(\bt\mu_{eff} ) \;  .
\ee
Using again the modified Bose function (\ref{19}) and following the same analysis
as in Sec. 3, we come to the conclusion that the system is stable for $d>2$, when
$T>T_c$. The difference is that now instead of mass $m$, there is the effective mass
$m^*$ and at the critical temperature, we have
$$
 \mu_{eff} = 0 \; , \qquad \mu = 2\rho\Phi_0 \qquad ( T = T_c ) \;  .
$$
So that the critical temperature becomes
\be
\label{94}
 T_c = \frac{2\pi}{m^*} \left[ \; \frac{\rho}{g_{d/2}(1) } \; \right]^{2/d} \;  .
\ee

As an example, let us consider the realistic three-dimensional case. Using the
Robinson representation (see details in review \cite{Yukalov_18}), we can find the
behaviour of the effective chemical potential at high temperatures
\be
\label{95}
\mu_{eff} = T \ln\left( \rho\lbd_T^3 \right) \qquad ( T \gg T_c )
\ee
and at the temperature approaching the critical point from above,
\be
\label{96}
 \mu_{eff} \simeq - T \; \frac{\zeta^2(3/2)}{4\pi} \; \left[ 1 -
\left( \frac{T_c}{T}\right)^{3/2} \right ]^2 \;  .
\ee
Then the isothermal compressibility
\be
\label{97}
\kappa_T = \frac{g_{1/2}(z)}{\rho^2 T \lbd^3_T}
\ee
at high temperatures is
\be
\label{98}
 \kappa_T \simeq \frac{1}{\rho T}  \qquad ( T \gg T_c ) \;  ,
\ee
while close to the critical point, it is
\be
\label{99}
\kappa_T \simeq \frac{0.921}{\rho T}  \left[ 1 -
\left( \frac{T_c}{T}\right)^{3/2} \right ]^{-1} \;   .
\ee
Respectively, particle fluctuations, described by the relative variance
\be
\label{100}
\frac{{\rm var}(\hat N)}{N} =  \rho T \kappa_T =
\frac{g_{1/2}(z)}{\rho\lbd_T^3} \; ,
\ee
at high temperatures behave as
\be
\label{101}
\frac{{\rm var}(\hat N)}{N} \; \simeq \; 1 \qquad ( T \gg T_c )
\ee
and close to the critical point, we get
\be
\label{102}
  \frac{{\rm var}(\hat N)}{N} \; \simeq \; 0.921 \left[ 1 -
\left( \frac{T_c}{T}\right)^{3/2} \right ]^{-1} \; .
\ee

Outside of the critical temperature itself, particle fluctuations are thermodynamically
normal. The divergence of the compressibility at the critical point signifies a
second-order phase transition. At the point of the phase transition, the system is not
stable and the fluctuations do not need to be finite.

\section{Interacting Bose System below Condensation Temperature}

In Sec. 3, it is proved that the ideal Bose gas, confined in a box, is stable below
the condensation temperature only for $d > 4$. In the present section, we show that
interactions stabilize the system making it stable already for $d=3$.

\subsection{Self-Consistent Approach}

For describing a Bose system with Bose-Einstein condensate, we employ the
self-consistent approach
\cite{Yukalov_16,Yukalov_17,Yukalov_18,Yukalov_22,Yukalov_32,Yukalov_33} providing
a gapless spectrum, correct thermodynamics, the validity of all conservation laws,
and good agreement with Monte Carlo simulations and experimental data.

The energy Hamiltonian has the form
$$
\hat H = \int \hat\psi^\dgr(\br) \left( -\; \frac{\nabla^2}{2m}
\right) \hat \psi(\br) \; d\br \; +
$$
\be
\label{103}
+ \;
\frac{1}{2} \int \hat\psi^\dgr(\br) \hat\psi^\dgr(\br')
\Phi(\br-\br') \hat \psi(\br') \hat\psi(\br)  \; d\br d\br' \;   .
\ee
The genuine Bose-Einstein condensation necessarily requires global gauge symmetry
breaking \cite{Lieb_6,Yukalov_9,Yukalov_17,Yukalov_18}. Finite systems, strictly
speaking, do not exhibit this symmetry breaking. However, a system with a large number
of particles $N \gg 1$ enjoys asymptotic symmetry breaking \cite{Birman_34} in the sense
that the system properties asymptotically, with respect to $N$, are close to the system
with broken symmetry. The global gauge symmetry can be broken by the Bogolubov shift
\cite{Bogolubov_26,Bogolubov_27,Bogolubov_28}
\be
\label{104}
 \hat \psi(\br) = \eta(\br) + \psi_1(\br) \;  ,
\ee
in which the condensate function $\eta({\bf r})$ and the operator of uncondensed
particles $\psi_1({\bf r})$ are mutually orthogonal,
\be
\label{105}
  \int \eta^*(\br) \psi_1(\br) \; d\br = 0
\ee
and the operator of uncondensed particles satisfies the condition
\be
\label{106}
 \lgl \; \psi_1(\br) \; \rgl = 0 \;  .
\ee

The number of condensed particles is
\be
\label{107}
N_0 = \int |\; \eta(\br)\; |^2 \; d\br \;   ,
\ee
while the number of uncondensed particles is given by the average
\be
\label{108}
 N_1 = \lgl \; \hat N_1 \; \rgl \; , \qquad
\hat N_1 = \int \psi_1^\dgr(\br) \psi_1(\br) \; d\br \;  .
\ee

The grand Hamiltonian reads as
\be
\label{109}
 H = \hat H - \mu_0 N_0 - \mu_1 \hat N_1 - \hat \Lbd \; ,
\ee
where
$$
\hat \Lbd = \int \left[ \lbd(\br) \psi_1^\dgr(\br) +
\lbd^*(\br)\psi_1(\br) \right] \; d\br
$$
and $\mu_0$, $\mu_1$, and $\lambda({\bf r})$ are Lagrange multipliers guaranteeing
the validity of normalizations (\ref{107}) and (\ref{108}), as well as condition
(\ref{106}).

The evolution equation for the condensate function can be written as
\be
\label{110}
 i\; \frac{\prt}{\prt t} \; \eta(\br,t) = \left\lgl \;
\frac{\dlt H}{\dlt\eta^*(\br,t) } \; \right\rgl
\ee
and the equation for the operator of uncondensed particles as
\be
\label{111}
  i\; \frac{\prt}{\prt t} \; \psi_1(\br,t) =
\frac{\dlt H}{\dlt\psi_1^\dgr(\br,t) } \; .
\ee

Keeping in mind, as usual, the periodic continuation of the box, we expand the field
operators in plane waves, as in (\ref{13}), and assume the existence of the Fourier
representation for the interaction potential
\be
\label{112}
 \Phi_k = \int \Phi(\br) e^{-i\bk\cdot\br}\; d\br \; , \qquad
\Phi(\br) = \frac{1}{V} \sum_k \Phi_k e^{i\bk\cdot\br} \;  .
\ee
Then we get the normal density matrix
\be
\label{113}
\rho_1(\br,\br') = \lgl \; \psi_1^\dgr(\br')\psi_1(\br) \; \rgl =
\frac{1}{V} \sum_{k\neq 0} n_k e^{i\bk\cdot(\br-\br')}
\ee
and the anomalous matrix
\be
\label{114}
\sgm_1(\br,\br') = \lgl \; \psi_1(\br')\psi_1(\br) \; \rgl =
\frac{1}{V} \sum_{k\neq 0} \sgm_k e^{i\bk\cdot(\br-\br')}   ,
\ee
in which
\be
\label{115}
 n_k \equiv \lgl \; a^\dgr_k a_k \; \rgl \; , \qquad
\sgm_k \equiv \lgl \; a_k a_{-k} \; \rgl \;  .
\ee

The condensate function $\eta({\bf r}) = \eta$ defines the condensate density
\be
\label{116}
 \rho_0 \equiv \frac{N_0}{V} = |\; \eta \; |^2 \;  .
\ee
The density of uncondensed particles is
\be
\label{117}
 \rho_1 \equiv \frac{N_1}{V} = \rho_1(\br,\br) = \frac{1}{V} \sum_k n_k \;  .
\ee
The diagonal anomalous matrix gives the anomalous average
\be
\label{118}
 \sgm_1 \equiv  \sgm_1(\br,\br) = \frac{1}{V} \sum_k \sgm_k \;  .
\ee
The average density of particles is the sum
\be
\label{119}
\rho \equiv \frac{N}{V} = \rho_0 + \rho_1 \; .
\ee

Then we use the Hartree-Fock-Bogolubov approximation and accomplish the Bogolubov
canonical transformation
$$
 a_k = u_k b_k + v^*_{-k} b^\dgr_{-k} \; , \qquad
b_k = u_k^* a_k - v_k^* a_{-k}^\dgr \;  ,
$$
where $u_k$ and $v_k$ are chosen so that to diagonalize the Hamiltonian. In that
way, we obtain the diagonalized Hamiltonian
\be
\label{120}
 H_B = E_B + \sum_k \ep_k b_k^\dgr b_k \;  ,
\ee
in which
$$
E_B = -\; \frac{1}{2} \; N \rho \Phi_0 \; - \; \rho_0 \sum_p ( n_p + \sgm_p) \Phi_p \;
- \; \frac{1}{2V} \sum_{kp} ( n_k n_p + \sgm_k \sgm_p ) \Phi_{k+p} \; + \;
\frac{1}{2} \sum_k (\ep_k - \om_k ) \; ,
$$
the particle spectrum is
\be
\label{121}
 \ep_k = \sqrt{\om_k^2 - \Dlt_k^2 } \;  ,
\ee
and where
$$
\om_k = \frac{k^2}{2m} + \Dlt + \rho_0 ( \Phi_k - \Phi_0 ) +
\frac{1}{V} \sum_p n_p ( \Phi_{k+p} - \Phi_p ) \; ,
$$
\be
\label{122}
 \Dlt_k = \rho_0 \Phi_k + \frac{1}{V} \sum_p \sgm_p \Phi_{k+p} \; , \qquad
 \Dlt \equiv \lim_{k\ra 0} \Dlt_k =
\rho_0 \Phi_0 + \frac{1}{V} \sum_p \sgm_p \Phi_p \; .
\ee

For the expressions in (\ref{115}), we find
\be
\label{123}
 n_k = \frac{\om_k}{2\ep_k} \; \coth\left( \frac{\ep_k}{2T}\right) - \;
\frac{1}{2} \; ,       \qquad
\sgm_k = -\;\frac{\Dlt_k}{2\ep_k} \; \coth\left( \frac{\ep_k}{2T}\right) \; .
\ee
And the chemical potentials are
\be
\label{124}
 \mu_0 = \rho \Phi_0 + \frac{1}{V} \sum_k ( n_k + \sgm_k ) \Phi_k \; , \qquad
\mu_1 = \rho \Phi_0 + \frac{1}{V} \sum_k ( n_k - \sgm_k ) \Phi_k \; .
\ee

In the long-wave limit, we can use the expansion
$$
\Phi_{k+p} \; \simeq \; \Phi_p + \frac{k^2}{2} \; \Phi_p'' \qquad ( k \ra 0 ) \;   ,
$$
where
$$
 \Phi_p'' \; \equiv \; \frac{\prt^2\Phi_p}{\prt p^2} \;  .
$$
Then spectrum (\ref{121}) becomes of the phonon type
\be
\label{125}
 \ep_k \; \simeq \; c k \qquad ( k \ra 0 ) \;  ,
\ee
with the sound velocity
\be
\label{126}
 c = \sqrt{ \frac{\Dlt}{m_{eff} }  }
\ee
and with the notation for the effective mass
\be
\label{127}
 m_{eff} \equiv \frac{m}{1 + \frac{m}{V} \sum_p(n_p - \sgm_p)\Phi_p''} \;  .
\ee
Actually, expression (\ref{126}), that can be written as
$$
 m_{eff} c^2 = \Dlt \; ,
$$
is the equation
\be
\label{128}
\frac{mc^2}{1 + \frac{m}{V} \sum_p(n_p - \sgm_p)\Phi_p'' } =
\rho_0 \Phi_0 + \frac{1}{V} \sum_p \; \sgm_p \Phi_p \; ,
\ee
defining the sound velocity $c$.

To simplify the consideration, we can resort to approximation (\ref{86}), similarly
to which we can write
\be
\label{129}
\sum_p \sgm_p \Phi_{k+p} \; \cong \; \Phi_k \sum_p \sgm_p \;   .
\ee
This gives
$$
\frac{1}{V} \sum_p ( n_p - \sgm_p) \Phi_p'' = (\rho_1 -\sgm_1) \Phi_0'' \;  ,
$$
where
$$
 \Phi_0'' = \lim_{p\ra 0} \Phi_p'' = - \;
\frac{4\pi}{3} \int_0^\infty \Phi(\br) r^4 \; d\br \;  .
$$
In view of the notation for the effective interaction radius (\ref{88}), we get
$$
\Phi_0'' = - \; \frac{1}{3} \; \Phi_0 r_{eff}^2 \; .
$$
Then the effective mass (\ref{127}) acquires the form
\be
\label{130}
m_{eff} = \frac{m}{1+(\sgm_1-\rho_1)\Phi_0 m r^2_{eff}/3} \; .
\ee

In approximations (\ref{86}) and (\ref{129}), the chemical potentials (\ref{124})
become
\be
\label{131}
 \mu_0 = \rho \Phi_0 + (\rho_1 + \sgm_1) \Phi_0 \; , \qquad
 \mu_1 = \rho \Phi_0 + (\rho_1 - \sgm_1) \Phi_0 \; .
\ee
Also we have
\be
\label{132}
 \om_k = \frac{k^2}{2m} + \Dlt + \rho (\Phi_k - \Phi_0 ) \; \qquad
\Dlt_k = (\rho_0 + \sgm_1) \Phi_k \; , \qquad
\Dlt = (\rho_0 + \sgm_1) \Phi_0 \;  .
\ee
The spectrum (\ref{121}) can be written as
\be
\label{133}
 \ep_k^2 = \left[ \frac{k^2}{2m} +
( \rho_1 - \sgm_1)( \Phi_k - \Phi_0) \right]
\left[ \frac{k^2}{2m} + \rho (\Phi_k - \Phi_0 ) +
( \rho_0 + \sgm_1)( \Phi_k + \Phi_0) \right] \;  .
\ee
The density of uncondensed particles is
\be
\label{134}
\rho_1 = \int \left[ \frac{\om_k}{2\ep_k} \;
\coth\left( \frac{\ep_k}{2T} \right) - \;
\frac{1}{2} \right] \; \frac{d\bk}{(2\pi)^3} \;   .
\ee
The anomalous average (\ref{118}) can be represented in the form
\be
\label{135}
\sgm_1 = - \int \frac{\Dlt_k}{2\ep_k} \; \frac{d\bk}{(2\pi)^3} \; - \;
\int \frac{\Dlt_k}{2\ep_k} \; \left[ \coth\left( \frac{\ep_k}{2T} \right) - 1
\right] \; \frac{d\bk}{(2\pi)^3} \;   .
\ee
When the first term here diverges, which happens for the local interaction, we can use
dimensional regularization \cite{Yukalov_18}.

\subsection{Particle Fluctuations}

The number-of-particle variance can be found by involving the formula
\be
\label{136}
 \frac{{\rm var}(\hat N)}{N} = 1 + \rho \int [ g(\br) - 1 ] \; d\br \;  ,
\ee
in which
\be
\label{137}
g(\br_{12}) = \frac{1}{g^2} \; \lgl \; \hat\psi^\dgr(\br_1) \hat\psi^\dgr(\br_2)
\hat\psi(\br_2) \hat\psi(\br_1) \; \rgl
\ee
is the pair correlation function, with $\br_{12}\equiv\br_1-\br_2$.

Accomplishing the Bogolubov shift (\ref{104}), we use the Hartree-Fock-Bogolubov
(HFB) decoupling for the expressions containing the operators $\psi_1$. Since
mathematically the HFB approximation is of second order with respect to the products
of the operators $\psi_1$, it is necessary to leave in the pair correlation function
only the terms of second order with respect to these operators
\cite{Yukalov_3,Yukalov_16,Yukalov_17,Yukalov_18,Yukalov_21,Yukalov_22}. As a result,
we obtain
\be
\label{138}
\int [ g(\br) - 1 ] \; d\br = \frac{2}{\rho} \; \lim_{k\ra 0} \; (n_k + \sgm_k) \; .
\ee
In this way, for the relative variance, we find
\be
\label{139}
  \frac{{\rm var}(\hat N)}{N} = 1 + 2 \lim_{k\ra 0} \; (n_k + \sgm_k) \;  .
\ee
For small $k$, when $\varepsilon_k$ tends to zero, we have
$$
n_k \; \simeq \; \frac{T\Dlt_k}{\ep_k^2} \; + \; \frac{\Dlt_k}{12T} \; + \;
\frac{T}{2\Dlt_k} \; - \; \frac{1}{2} \; + \; \left( \frac{\Dlt_k}{3T}\; - \;
\frac{T}{\Dlt_k} \; - \; \frac{\Dlt_k^3}{90T^3} \right)\;
\frac{\ep^2_k}{8\Dlt_k^2} \; ,
$$
\be
\label{140}
\sgm_k \; \simeq \; -\; \frac{T\Dlt_k}{\ep_k^2}\; - \; \frac{\Dlt_k}{12T} \; + \;
\frac{\Dlt_k\ep_k^2}{720T^3} \qquad ( \ep_k \ra 0 )  \; .
\ee
Therefore
$$
\lim_{k\ra 0} \; (n_k + \sgm_k) =
\frac{1}{2} \left( \frac{T}{\Dlt} \; - \; 1 \right) \; ,
$$
with
$$
\Dlt = m_{eff} c^2 = ( \rho_0 + \sgm_1) \Phi_0 \;  .
$$
Thus we come to the expression
\be
\label{141}
\frac{{\rm var}(\hat N)}{N} = \frac{T}{m_{eff}c^2} \; .
\ee
Respectively, the compressibility is
\be
\label{142}
  \kappa_T = \frac{{\rm var}(\hat N)}{N\rho T} = \frac{1}{\rho m_{eff}c^2} \; .
\ee
Taking into account formula (\ref{126}) leads to the variance
\be
\label{143}
 \frac{{\rm var}(\hat N)}{N} = \frac{T}{(\rho_0+\sgm_1)\Phi_0} \;  .
\ee

Note that expression (\ref{143}) is valid at zero temperature as well. This is easy
to check considering quantities (\ref{123}) at zero temperature,
$$
n_k = \frac{\sqrt{\ep_k^2+\Dlt^2_k}}{2\ep_k} \; - \; \frac{1}{2} \; , \qquad
\sgm_k = -\; \frac{\Dlt_k}{2\ep_k} \qquad ( T = 0 )  \; .
$$
From here, in the long-wave limit, we have
$$
n_k \; \simeq \; \frac{\Dlt_k}{2\ep_k} \; + \; \frac{\ep_k}{4\Dlt_k} \; - \;
\frac{1}{2}  \qquad ( \ep_k \ra 0 \; , ~ T = 0 ) \;   .
$$
Hence
$$
 \lim_{k\ra 0}\; (n_k + \sgm_k) = - \; \frac{1}{2}   \qquad ( T = 0 )
$$
and
$$
 \frac{{\rm var}(\hat N)}{N} = 0 \qquad ( T = 0 ) \; .
$$

The above result for the relative variance (\ref{143}) can be generalized for
nonuniform systems \cite{Yukalov_35} by involving the local-density approximation,
which yields
\be
\label{144}
\frac{{\rm var}(\hat N)}{N} = \frac{T}{N}
\int \frac{\rho(\br)}{\Dlt(\br)} \; d\br \;   ,
\ee
where
\be
\label{145}
 \Dlt(\br) = [ \; \rho_0(\br) + \sgm_1(\br)\; ] \Phi_0 \;  .
\ee

Particle fluctuations in a three-dimensional Bose-condensed system of interacting
particles are thermodynamically normal in both the cases, when particles are in
a box or in a nonuniform external potential.

\section{Conclusion}

Particle fluctuations in Bose systems are studied. Investigating the behavior of these
fluctuations is important because they are directly connected with isothermal compressibility
and define the system stability with respect to pressure variations. Thermodynamically
anomalous fluctuations signify system instability. While thermodynamically normal
fluctuations mean that the equilibrium system is stable. The obtained results are
as follows.

Ideal Bose gas confined in a rectangular box is stable, depending on temperature, in spatial
dimensions
$$
d \; > \; 2 \qquad ( T > T_c) \; ,
$$
$$
 d \; > \; 4 \qquad ( T < T_c) \;  .
$$

The stability of ideal Bose gas in a power-law trap depends on the confining
dimension
$$
 D \equiv d + \sum_{\al=1}^d \frac{2}{n_\al} \;  .
$$
This gas is stable for the confining dimensions
$$
D \; > \; 2 \qquad ( T > T_c) \; ,
$$
$$
D \; > \; 4 \qquad ( T < T_c) \;   .
$$

Interactions stabilize Bose-condensed systems, so that an interacting system with
Bose-Einstein condensate becomes stable at $d = 3$ for either a system in a box or
in an external potential.

Nonlocal interactions with a stronger strength or with a larger interaction radius
increase the effective mass, hence diminish the condensation temperature.

In conclusion, it is worth stressing that the knowledge of particle or charge fluctuations 
not only signifies the system stability or instability, but also gives more detailed 
information on other system properties. For example, there exists an interesting relation 
between fluctuations and von Neumann entanglement entropy 
\cite{Song_38,Song_39,Song_40,Rachel_41,Petrescul_42}.   

\section*{Acknowledgments}

The author is grateful to E.P. Yukalova for useful discussions.

\end{document}